\documentclass[12pt]{article}
\usepackage{cmap}
\usepackage{cyrillic,amssymb,revsymb}
\newcommand{\cyrrm}{\fontencoding{OT2}\fontfamily{wncyr}\selectfont\textcyrup}
\newcommand{\cyrit}{\fontencoding{OT2}\fontfamily{wncyr}\selectfont\textcyrit}

\usepackage[square,comma,numbers,compress]{natbib}

\usepackage[linktocpage=true,dvips,bookmarksnumbered=true,breaklinks=true,pagebackref=false,colorlinks=true,urlcolor=blue,citecolor=red,linkcolor=magenta]{hyperref}
\usepackage{hypernat}
\usepackage{url}
\usepackage{breakurl}

\begin{document}
\bibliographystyle{plainnat}
\def\refname{\Large References\phantomsection
\addcontentsline{toc}{section}{References}

\noindent
{\rm\small 
\begin{minipage}[t]{15cm}
[For references in Cyrillic letters, 
we apply the (new) {\it Mathematical Reviews} transliteration
(transcription) scheme
(to be found at the end of index issues of {\it Mathematical Reviews}).]
\end{minipage}}\\[-0.3cm]}

\ \\[3cm]

\renewcommand{\thefootnote}{\fnsymbol{footnote}}
\begin{center}
{\LARGE\baselineskip0.9cm
Entanglement capabilities of the spin representation 
of (3+1)D-conformal transformations\\[1.5cm]}

{\large 
K. Scharnhorst\footnote[2]{E-mail: {\tt k.scharnhorst@vu.nl}}
}\\[0.3cm]

{\small 
Vrije Universiteit Amsterdam,
Faculty of Sciences, Department of Physics and Astronomy,
De Boelelaan 1081, 1081 HV Amsterdam, The Netherlands}\\[1.5cm]

\begin{abstract}
Relying on a mathematical analogy of the pure states of the 
two-qubit system of quantum information theory with
four-component spinors we introduce the concept of 
the intrinsic entanglement of spinors. To explore its physical sense
we study the entanglement capabilities of the spin representation
of (pseudo-) conformal
transformations in (3+1)-dimensional Minkowski space-time.
We find that only those tensor product structures can
sensibly be introduced in spinor space
for which a given spinor is not entangled.
\end{abstract}

\end{center}

\renewcommand{\thefootnote}{\arabic{footnote}}
\thispagestyle{empty}

\newpage

\section{Introduction}

Fermions are an essential part of physical reality.
In the realm of relativistic quantum mechanics and
quantum field theory in 3+1-dimensional Minkowski space-time
they are mainly described by means of Dirac spinors
-- elements of a four-component complex vector space
which are transforming under
the spin representation of the Poincar\'e group
(inhomogeneous Lorentz group) of space-time.
The Dirac equation
\begin{eqnarray}
\label{ec1}
(i\slash\hspace{-0.2cm}\partial - m) \Psi (x)
&=&(i\gamma_\mu\partial^\mu- m) \Psi (x)\ =\ 0
\end{eqnarray}
is form-invariant under the Poincar\'e group
[Here, $\gamma_\mu$ denote the standard 
$4\times 4$ gamma matrices,
$\gamma_\mu\partial^\mu = g^{\mu\nu}\gamma_\mu\partial_\nu$,
$\mu, \nu = 1\ldots 4$,
with $g = {\rm diag}(-1,-1,-1,1)$.].
For massless fermions ($m = 0$), the Dirac equation is even
form-invariant under the 
larger (15-dimensional) conformal group
which is obtained by extending the 10-dimensional
Poincar\'e group by special (pseudo-) conformal transformations
and dilatations. For massless fermions, the Dirac 
equation separates into two equations for spinors 
\begin{eqnarray}
\label{ec1b}
\Psi_L &=&\frac{1}{2}(\openone -i\gamma_5)\Psi ,\\[0.3cm]
\label{ec1c}
\Psi_R &=&\frac{1}{2}(\openone +i\gamma_5)\Psi ,
\end{eqnarray}
$\Psi = \Psi_L\oplus \Psi_R$,
of different chirality ($\gamma_5 = \gamma_1\gamma_2\gamma_3\gamma_4$).
If $\gamma_5$ is chosen to be diagonal $\Psi_L$ and $\Psi_R$
are two-component objects (Weyl spinors). However, also in the massive 
case ($m \not= 0$) the analysis of $\Psi$ is often pursued by
representing it as the {\it direct sum} of
2 two-component objects, e.g., the small and large 
components of the Dirac spinor $\Psi$ (see, e.g., \citep{schw1}, 
Chap.\ 9). The mathematically possible 
representation of a Dirac spinor in 
terms of {\it direct (tensor) products} \cite{foot1}
of 2 two-component objects has
been studied so far by Tokuoka (\citep{toku1}, Part II, p.\ 161)
in connection with the wave matrix formalism
of G\"ursey (\citep{gurs1}, \citep{gurs2}, Sec.\ 3, p.\ 997),
and by Uschersohn \citep{usch1} (see pp.\ 25-31) only.

The theory of quantum information and computation has received
considerable attention in recent years. Due to the fundamental
significance of quantum information theory, in the past decade 
the application of its principles and viewpoints within relativistic
quantum mechanics and quantum field theory has begun to be explored
\citep{pere1}. One of the most fundamental and most studied models of 
quantum information theory is the two-qubit (two-spin-$\frac{1}{2}$)
system (see, e.g., \citep{zhan1}). 
The pure states of the two-qubit system are described in terms
of vectors of a four-dimensional Hermitian (complex) 
vector space $\mathbb{C}_4$
and quantum operations in it correspond to elements of the group
$SU(4)$. Due to the mathematical similarities it is now tempting 
to explore the connection between one 
of the most fundamental models of relativistic quantum mechanics --
four-component Dirac spinors -- and one of the most fundamental
models of quantum information theory -- the two-qubit system.
The present article is devoted to certain aspects of this problem.
The main result of our analysis will be that only those tensor 
product structures can sensibly be introduced in spinor space 
for which a given spinor is not entangled.\\

The transition from the pure states of a
two-qubit system to four-component Dirac
spinors we are studying can most easily be understood as a two 
step process proceeding over the intermediate stage of Euclidean
spinors. The first step from the two-qubit system to 
four-component Euclidean spinors (spinors of $\mathbb{R}_6$)
consists in a re-interpretation of the mathematical elements
used to describe the two-qubit system. Unitary $SU(4)$
transformations arising in two-qubit systems from the (unitary) action of a
Hamiltonian onto the two $SU(2)$ spins can be re-interpreted
as the (unitary) spinor representation of special orthogonal 
transformations $SO(6;\mathbb{R})$ in some Euclidean vector space 
$\mathbb{R}_6$ which we will denote by the term `base space'. 
Consequently, in this re-interpretation each action of
a two-spin-$\frac{1}{2}$ Hamiltonian generating a $SU(4) = Spin(6)$
transformation stands in correspondence to a 
spatial transformation in some base space $\mathbb{R}_6$. This simply amounts
to a geometrical (spatial) re-interpretation of the unitary transformations 
occurring in any two-qubit system. The second step, from Euclidean spinors 
to Minkowski space spinors, relies on the 
isomorphism $SL(4;\mathbb{C})/\mathbb{Z}_2\simeq SO(6;\mathbb{C})$
\cite{foot2}. The transition from Euclidean to Minkowski 
space is a standard procedure, 
e.g.\ in quantum field theory, by re-interpreting one spatial 
coordinate of the Euclidean base
space $\mathbb{R}_6$ as time coordinate $x_6$ and to go over 
to the time coordinate $t$ of Minkowski space $\mathbb{R}_{5,1}$
by means of a Wick rotation $x_6 = i t$
(The real-world four-dimensional Minkowski space $\mathbb{R}_{3,1}$
can of course be thought of to be embedded in $\mathbb{R}_{5,1}$.). 
The latter transition 
can be understood as an analytic continuation from $\mathbb{R}_6$ 
to $\mathbb{R}_{5,1}$, both spaces being embedded in some
six-dimensional complex vector space $V_\mathbb{C}$. 
The Minkowski space analogues of special orthogonal 
transformations $SO(6;\mathbb{R})$ in the base space 
$\mathbb{R}_6$ are then Lorentz 
transformations in $\mathbb{R}_{5,1}$, and the transition between 
the corresponding spinor representations, from $Spin(6) = SU(4)$ to 
$Spin(5,1) = SU^{\displaystyle\ast}(4) = SL(2,\mathbb{H})$ \cite{foot3}, 
can also be understood as an analytic 
continuation in the framework of the 
$SL(4;\mathbb{C})/\mathbb{Z}_2\simeq SO(6;\mathbb{C})$ isomorphism
whereby both spinor groups can be thought of to be embedded 
in the group $SL(4,\mathbb{C})$ as subgroups. Applying the method
of analytic continuation to some further $V_\mathbb{C}$ space coordinate
brings us to the consideration of the group of pseudo-orthogonal transformations
$SO(4,2;\mathbb{R})$ in some base space 
$\mathbb{R}_{4,2}\subset V_\mathbb{C}$ with
its associated spin group $Spin(4,2) = SU(2,2)$ (again, as subgroups in the
framework of the $SL(4;\mathbb{C})/\mathbb{Z}_2\simeq SO(6;\mathbb{C})$
isomorphism). In $\mathbb{R}_{4,2}$, (pseudo-) 
conformal transformations of Minkowski
space $\mathbb{R}_{3,1}$ (embedded in $\mathbb{R}_{4,2}$) 
can linearly be represented by means of 
pseudo-orthogonal transformations $SO(4,2;\mathbb{R})$ and 
the group of $SU(2,2)$ transformations is the related spin representation
(cf., e.g., \cite{kast1}). Formally, $SU(2,2)$ transformations can be 
considered as being generated by the action of Hamiltonians 
of two $SU(1,1)$ (quasi-) spins.
Qualitatively, the differences between the pure states of two-qubit systems and
Dirac spinors can be summarized as follows
(Here, the $4\times 4$ matrix $\eta$ defines the quadratic form 
in $\mathbb{C}_{2,2}$.):

\vspace{1cm}

\begin{tabular}[h]{lcc}
&two-qubit system & Dirac spinors\\[0.3cm]
\cline{1-3}
&&\\
base space & $\mathbb{R}_6$&$\mathbb{R}_{4,2}$\\
&&($\mathbb{R}_{3,1}$ embedded)\\[0.3cm]
spin(or) space & $\mathbb{C}_4$&$\mathbb{C}_{2,2}$\\[0.3cm]
spin group & $SU(4) = Spin(6)$&$SU(2,2) = Spin(4,2)$\\[0.3cm]
generators
&Hermitian:
&pseudo-Hermitian:\\
(``spin Hamiltonians'')& $H = H^\dagger$&$\eta H \eta^{-1} = H^\dagger$\\
of the spin group& &
\end{tabular}
 
\vspace{1cm}

\noindent
There is no significant conceptual difference concerning the
introduction of tensor product structures in the respective 
spin spaces $\mathbb{C}_4$ and $\mathbb{C}_{2,2}$. However, as we
will see in sec.\ \ref{secschmidt} the extension of the Schmidt 
decomposition from tensor product structures in $\mathbb{C}_4$ to
$\mathbb{C}_{2,2}$ meets certain difficulties.\\

\section{Spinor space}

In difference to pure states of the two-qubit system,
Dirac spinors in (3+1)-dimen\-sional Minkowski space
live in an indefinite complex vector space
$\mathbb{C}_{2,2}$ and the spin representation of the (pseudo-) conformal group
of Minkowski space corresponds to the group $SU(2,2)$. This difference 
needs to be taken care of. To prepare ourselves for this task
we shortly recall here the definition of the (indefinite) scalar product for 
Dirac spinors. The adjoint spinor $\bar{\Psi}$ to $\Psi$ is
defined as usual by means of the relation
\begin{eqnarray}
\label{ec2}
\bar{\Psi}&=&\Psi^\dagger \gamma_4
\end{eqnarray}
and the [Lorentz/Poincar\'e/(pseudo-) conformally invariant] 
scalar product of two spinors $\Psi$ and $\Phi$ is given by
\begin{eqnarray}
\label{ec3}
\langle\Psi,\Phi\rangle&=&\bar{\Psi}\Phi\ =\ \Psi^\dagger\gamma_4\Phi\ .
\end{eqnarray}
The concrete shape of the scalar product depends on the choice of
the gamma matrix $\gamma_4$ ($= \eta$). 
We will choose $\gamma_4$ as non-entangled
(for the concept of operator entanglement see \citep{zana4}), i.e.,
as the tensor product of two $2\times 2$-matrices $\kappa_A$, $\kappa_B$:
\begin{eqnarray}
\label{ec4}
\gamma_4&=&\kappa_A\otimes\kappa_B\ =\ \eta\ .
\end{eqnarray}
For most of the discussion, we will not need to refer to any specific choice
of $\kappa_A$, $\kappa_B$, however, for any explicit calculation we
will rely on
\begin{eqnarray}
\label{ec7}
\kappa_A&=&\kappa_B\ =\ \sigma_3
\end{eqnarray}
where $\sigma_3 = diag(1,-1)$ is the standard third Pauli matrix.
For an entangled representation of the gamma matrices see, e.g., 
\cite{scha1}, p.\ 2872, eqs.\ (24)-(27).\\

The concept of entanglement assumes a central role in quantum
information and computation theory \citep{horo1}. 
It is well known that entanglement
(characterized by means of some entanglement measure, see, e.g., 
\citep{plen1}) is not an absolute characteristics of a quantum state
but depends on the tensor product structure (TPS) chosen in the
quantum state space \citep{zana1,zana2,hars1,torr1,hars2}. Therefore, in recent
years attention has been given to the quantification of the 
changes in entanglement brought
about by quantum operations (initially this has been 
discussed in \citep{zana3,dur1}). 
In the present article, we will be interested in the concept of the 
entanglement capability of a single (quantum) operation \citep{dur1}, 
not in the statistical concept of the entangling power \cite{zana3}
of (quantum) operations. As the physical significance
of a tensor product structure for Dirac spinors is unknown so far,
it seems to be a sensible choice to concentrate in a first analysis 
on possible changes in the entanglement --
the entanglement capability of spin transformations.
In view of the mathematical similarity between Dirac spinors and the 
space of pure states of the two-qubit system, in the following we 
will rely on the formalism developed in \citep{dur1} and study
the entanglement capabilities of infinitesimal (pseudo-) conformal
transformations in (3+1)-dimensional Minkowski space-time.\\

\section{\label{secschmidt}The generalized Schmidt decomposition}

We start the discussion by displaying the (generalized) Schmidt decomposition
(cf.\ eq.\ (2) in \citep{dur1}) of an arbitrary spinor $\Psi$.
At this stage, $\Psi$ is to us simply an element of the indefinite complex
vector space $\mathbb{C}_{2,2}$, and not necessarily depending on
Minkowski space-time coordinates or in any way related to the Dirac equation.
The (generalized) Schmidt decomposition of a given spinor $\Psi$ 
reads ($P\in\mathbb{R}$, $\frac{1}{2}\le P\le 1$)
\begin{eqnarray}
\label{ec5}
\Psi&=&\sqrt{P}\ \psi_A\otimes\psi_B\
+\ \sqrt{1-P}\ \psi_A^\perp\otimes\psi_B^\perp\ .
\end{eqnarray}
Here, we disregard null-spinors by setting 
$\vert\langle\Psi,\Psi\rangle\vert = 1$.
Furthermore, we set
$\vert\langle\psi_L,\psi_L\rangle\vert = 
\vert\langle\psi^\perp_L,\psi^\perp_L\rangle\vert = 1$, ($L = A,B$)
where $\psi_L$, $\psi_L^\perp$ span the two-dimensional space $\mathbb{K}_L$
the matrix $\kappa_L$ operates in 
(For solutions $\Psi$ of the Dirac equation,
the Schmidt coefficient $P$ may depend on space-time
coordinates or four-momenta in general.). In view of eq.\ (\ref{ec4}),
the scalar product of two non-entangled ($P = 1$) spinors
$\Psi = \psi_A\otimes\psi_B$, $\Phi = \phi_A\otimes\phi_B$, factorizes into
the scalar products of the 2 two-component spaces $\mathbb{K}_L$:
\begin{eqnarray}
\label{ec6}
\langle\Psi,\Phi\rangle&=&
\langle\psi_A,\phi_A\rangle_{\kappa_A}
\langle\psi_B,\phi_B\rangle_{\kappa_B}\ ,\\[0.3cm]
\langle\psi_L,\phi_L\rangle_{\kappa_L}&=&\psi_L^\dagger\kappa_L\phi_L\ .
\end{eqnarray}
The notation $\psi_L^\perp$ denotes vectors with
$\langle\psi_L,\psi_L^\perp\rangle_{\kappa_L} = 0$.\\

The generalized Schmidt decomposition (\ref{ec5}) requires some comment.
Depending on the chosen TPS, not for every spinor $\Psi$ such a 
generalized Schmidt decomposition may exist and in this article
we will consider only those spinors $\Psi$ for which (for a given TPS) 
an equation of the type (\ref{ec5}) can be found. The reason for 
this situation can be recognized from the tensor product representation
of $\gamma_4$ [eq.\ (\ref{ec4})] which entails that at least one of
the two-component spaces $\mathbb{K}_L$, $L = A, B$, must be 
equipped with an indefinite scalar product. 
For our choice, eq.\ (\ref{ec7}), we can resort to the hyperbolic singular 
value decomposition (SVD) \citep{sego1} (Theorem 3.4, p.\ 1268)
which under fairly general conditions grants us the existence of
a generalized Schmidt decomposition (by virtue of the close connection
between the Schmidt decomposition and the SVD, see, e.g., 
\citep{niel1}, Sec.\ 2.5, p.\ 109). It should be mentioned here 
that theorem 3.4 in \citep{sego1} provides us only with real 
coefficients in the (generalized) Schmidt decomposition. But taking
into account the fact that in our consideration 
the spaces $\mathbb{K}_A$, $\mathbb{K}_B$ are independent of each other
we can always choose the coefficients to be non-negative (by means
of a reflection). This 
is in agreement with insight coming from the generalized (left)
polar decomposition \citep{high1} (Theorem 3.9, p.\ 2173; 
for the relation of the SVD to the polar decomposition
see, e.g., \citep{niel1}, Sec.\ 2.1.10, p.\ 78).
Finally, we would like to comment on making a different choice than
eq.\ (\ref{ec7}). For illustration, let us choose the example
$\kappa_A = \sigma_3$, $\kappa_B = \openone_2$
($\openone_2$ is the $2\times 2$-unit matrix.).
Then, we would have to resort to
the generalized (left) polar decomposition \citep{high1} to derive
a generalized Schmidt decomposition (In our example the spaces
$\mathbb{K}_A$ and $\mathbb{K}_B$ are no longer isomorphic
because the scalar products related to them differ qualitatively.).
We find that for our example the conditions of the theorem 3.9 in
\citep{high1}, p.\ 2173, are not fulfilled and a generalized 
Schmidt decomposition is not at hand. Therefore, we will not further
consider the last example.\\

\section{\label{spin}Spin representation of conformal transformations}

The application of the principles and viewpoints of
quantum information theory to the domain of relativistic quantum 
physics has a fairly recent history only (see \citep{pere1}).
Most studies have been based on an unitary, infinite-dimensional 
representation of the Poincar\'e group 
to investigate relativistic aspects of the entanglement of two spins
(\citep{pere3} and follow-up
articles citing it, \citep{pere1}, Sec.\ IV, p.\ 105.
N.B.: There are no finite-dimensional
{\it unitary} representations of the Poincar\'e group because it is 
a non-compact group.).
In contrast, we will study here a non-unitary, finite-dimensional
(spin) representation of the Poincar\'e group \citep{hepn1,kast1}
which can be extended to a representation of the (pseudo-) conformal group of
(3+1)-dimensional Minkowski space. The generators of this 
representation read (\citep{kast1}, p.\ 411, eqs.\ (19a), (19b); the
two possible signs refer to two inequivalent representations;
$k,l = 1,2,3$):
\begin{eqnarray}
\label{ec8a}
M_{\mu\nu}&=&-\ M_{\nu\mu}\ =\ 
\frac{i}{4}\left(\gamma_\mu\gamma_\nu 
- \gamma_\nu\gamma_\mu\right)\ ,\\[0.3cm]
&&\ \ \ \ M_{k4}^\dagger\ =\ - M_{k4},\
M_{kl}^\dagger\ =\ M_{kl},\nonumber\\[0.3cm]
\label{ec8b}
D&=&\mp \frac{1}{2}\ \gamma_5\ =\ - D^\dagger\ ,\\[0.3cm]
\label{ec8c}
P_\mu&=&\pm \frac{1}{2}\gamma_\mu\left(\openone \mp i \gamma_5\right)\ ,
\ \ P_4^\dagger\ =\ K_4,\ P_k^\dagger\ =\ - K_k,\\[0.3cm] 
\label{ec8d}
K_\mu&=&\pm \frac{1}{2}\gamma_\mu\left(\openone \pm i \gamma_5\right)\ ,
\ \ K_4^\dagger\ =\ P_4,\ K_k^\dagger\ =\ - P_k.
\end{eqnarray}
Here, $M_{\mu\nu}$ denote the generators of Lorentz transformations,
$D$ the generator of dilatations, $P_\mu$ the generators of space-time
translations, and $K_\mu$ the generators of 
special (pseudo-) conformal transformations.
For the calculation of the Hermitian conjugates we have
used the relations [cf.\ our choice (\ref{ec7})]:
$\gamma_4 = \gamma_4^\dagger$, $\gamma_k = - \gamma_k^\dagger$.\\ 

\section{Entanglement changes}

We will be interested in the question
how the space-time symmetries of a (massless) fermion influence the
entanglement characteristics of a spinor describing it. Of course,
the understanding of the concept of entanglement applied here is
in various respect a generalized one and differs from the one 
conventionally applied in the analysis of two spatially separated
spins. We will denote the type of entanglement we have in mind by the term
{\it intrinsic entanglement}.
Primarily, we will be interested in the question if a
(pseudo-) conformal transformation leads to any change in the 
intrinsic entanglement of a spinor, consequently an 
infinitesimal approach is appropriate. We will proceed
as the authors of ref.\ \citep{dur1}. Suppose we have got an 
(generalized) entanglement measure $E$ (e.g., the entropy of 
entanglement) which only depends on the Schmidt coefficient $P$. 
The rate of the change of the entanglement $E$ under
a (space-time) transformation parameterized by some infinitesimal 
parameter $\tau\in\mathbb{R}$ then reads
\begin{eqnarray}
\label{ec9}
\Gamma(\tau)&=&\frac{dE}{d\tau}\ =\ \frac{dE}{dP}\ \frac{dP}{d\tau}\ . 
\end{eqnarray}
We will exclusively be interested in the question
if the quantity $\dot{P} = \frac{dP}{d\tau}$ vanishes or not [If so,
the intrinsic entanglement (measure) $E$ remains constant.]. 
To obtain an explicit 
expression for $\dot{P}$ for the spin representation of an 
infinitesimal space-time transformation
$\exp(- i H\tau) = (\openone_4 - i H\tau)$ 
it is useful to consider the reduced (relativistic)
density operator for the subspace (say) $A$:
\begin{eqnarray}
\label{ec10a}
\rho_A(\tau)&=&{\rm tr}_B\left[\rho(\tau)\right]\nonumber\\[0.3cm]
&=&\rho_A(0)\ 
-\ i\tau\ {\rm tr}_B\left[H\rho(0)- \rho(0) \gamma_4 H^\dagger
  \gamma_4\right]\ .\ \ \ 
\end{eqnarray}
$H$ denotes here any of the generators of 
(pseudo-) conformal transformations (\ref{ec8a})- (\ref{ec8d}) which are
not Hermitian in general, however (In this respect the situation differs
from the consideration in \citep{dur1}.).
Eq.\ (\ref{ec10a}) is derived by means of
\begin{eqnarray}
\label{ec10b}
\rho(\tau)&=&\frac{\Psi(\tau)\otimes\bar{\Psi}(\tau)}{
\langle\Psi,\Psi\rangle}
\ = \frac{\Psi(\tau)\otimes\Psi(\tau)^\dagger \gamma_4}{
\langle\Psi,\Psi\rangle}
\ ,\ \ \ \\[0.3cm]
\label{ec10c}
\Psi(\tau)&=&{\rm e}^{(- i H\tau)}\Psi\ .
\end{eqnarray}
The partial trace can be defined by means of the following equation. 
\begin{eqnarray}
\label{ec10d}
\rho_A(\tau)&=&{\rm tr}_B\left[\rho(\tau)\right]\nonumber\\[0.3cm]
&=&\frac{\langle\psi_B(\tau),\rho(\tau)\psi_B(\tau)\rangle_{\kappa_B}}{
\langle\psi_B(\tau),\psi_B(\tau)\rangle_{\kappa_B}}\ +\ 
\frac{\langle\psi_B^\perp(\tau),\rho(\tau)\psi_B^\perp(\tau)\rangle_{\kappa_B}}{
\langle\psi_B^\perp(\tau),\psi_B^\perp(\tau)\rangle_{\kappa_B}}
\end{eqnarray}
 $\rho(\tau)\psi_B^\perp$ denotes
-- somewhat symbolically -- a vector in the (spin) space which $\kappa_B$ is
operating in and which can be determined on the basis of the tensor 
product decomposition of the density operator $\rho(\tau)$ with respect to
the vector spaces  $\mathbb{K}_A$ and  $\mathbb{K}_B$.\\

In eq.\ (\ref{ec10a}) besides the operator (matrix) $H$ the 
object $\gamma_4 H^\dagger \gamma_4$ makes its appearance.  
For all the generators $H$ of (pseudo-) conformal transformations 
(\ref{ec8a})-(\ref{ec8d}) the equation
\begin{eqnarray}
\label{ec11}
\gamma_4 H^\dagger \gamma_4&=&H
\end{eqnarray}
applies.
Consequently, eq.\ (\ref{ec10a}) can be written in our non-Hermitian case
exactly as in ref.\ \cite{dur1} (above of eq.\ (5)) as
\begin{eqnarray}
\label{ec12}
\rho_A(\tau)&=&\rho_A(0)\ 
-\ i\tau\ {\rm tr}_B\left(\left[H,\rho(0)\right]\right)\ .
\end{eqnarray}
We would like to mention here that the relation (\ref{ec11}) 
is of the same type as that 
for pseudo-Hermitian Hamiltonians considered
in non-Hermitian quantum theory. For a discussion of 
aspects of non-Hermitian quantum information theory the reader is
referred to \citep{scol1,pati1}.\\

Relying on the eqs.\ (\ref{ec10b}), (\ref{ec10d}),
from eq.\ (\ref{ec5}) one finds the following 
representation for $\rho_A$
($\langle\psi_B(\tau),\psi_B(\tau)\rangle_{\kappa_B}
= \langle\psi_B,\psi_B\rangle_{\kappa_B}$,
$\langle\psi_B^\perp(\tau),\psi_B^\perp(\tau)\rangle_{\kappa_B}
= \langle\psi_B^\perp,\psi_B^\perp\rangle_{\kappa_B}$)
\begin{eqnarray}
\label{ec13}
\rho_A(\tau)&=&\frac{1}{\langle\Psi,\Psi\rangle}
\left(P\ \langle\psi_B,\psi_B\rangle_{\kappa_B}\
\psi_A(\tau)\otimes\psi_A^\dagger(\tau)\kappa_A\right.\nonumber\\[0.3cm]
&&\left.\ \ +\
(1-P)\  \langle\psi_B^\perp,\psi_B^\perp\rangle_{\kappa_B}\
\psi_A^\perp(\tau)\otimes\psi_A^{\perp\dagger}(\tau)\kappa_A\right)\ .\ \ \ 
\end{eqnarray}
We differentiate eq.\ (\ref{ec13}) now with respect to $\tau$ 
and take the scalar product\hfill\ \linebreak
$\langle\psi_A(\tau),\dot{\rho}_A(\tau)\psi_A(\tau)\rangle_{\kappa_A}$:
\begin{eqnarray}
\label{ec13b}
\langle\psi_A(\tau),\dot{\rho}_A(\tau)\psi_A(\tau)\rangle_{\kappa_A}
&=&\frac{\dot{P}}{\langle\Psi,\Psi\rangle}\ 
\langle\psi_B,\psi_B\rangle_{\kappa_B}\
\left(\langle\psi_A,\psi_A\rangle_{\kappa_A}\right)^2\ . \ \ \ 
\end{eqnarray}
To arrive at eq.\ (\ref{ec13b}) we have taken into account
the relation (derived from\hfill\ \linebreak
$\langle\psi_A(\tau),\psi_A(\tau)\rangle_{\kappa_A}
= \langle\psi_A(0),\psi_A(0)\rangle_{\kappa_A}$)
\begin{eqnarray}
\label{ec13c}
\langle\dot{\psi}_A(\tau),\psi_A(\tau)\rangle_{\kappa_A} +
\langle\psi_A(\tau),\dot{\psi}_A(\tau)\rangle_{\kappa_A}&=&0\ . \ \ \ 
\end{eqnarray}
Note, that the term proportional to $(1-P)$ present in eq.\ (\ref{ec13})
drops out in eq.\ (\ref{ec13b}) in view of the orthogonality conditions
$\langle\psi_A^\perp,\psi_A\rangle_{\kappa_A} = 
\langle\psi_A,\psi_A^\perp\rangle_{\kappa_A} = 0$.
Taking into account the normalization conditions
$\vert\langle\Psi,\Psi\rangle\vert = 1$,
$\vert\langle\psi_L,\psi_L\rangle\vert = 
\vert\langle\psi^\perp_L,\psi^\perp_L\rangle\vert = 1$, ($L = A,B$)
mentioned below from eq.\ (\ref{ec5}), we find
\begin{eqnarray}
\label{ec14}
\dot{P}&=&\dot{P}(\tau)\ =\ 
\frac{\langle\psi_A(\tau),\dot{\rho}_A(\tau)\psi_A(\tau)\rangle_{\kappa_A}}{
\langle\psi_A,\psi_A\rangle_{\kappa_A}}\ .\ \ \
\end{eqnarray}
Using eq.\ (\ref{ec12}), we then obtain (for $\tau = 0$)
\begin{eqnarray}
\label{ec14b}
\dot{P}&=&-i\ 
\frac{\langle\psi_A,{\rm tr}_B\left(\left[H,\rho(0)\right]\right)
\psi_A\rangle_{\kappa_A}}{
\langle\psi_A,\psi_A\rangle_{\kappa_A}}\ .\ \ \
\end{eqnarray}
Relying on eq.\ (\ref{ec10b}) (for $\tau = 0$),  we can now insert
into $\rho(0)$ the generalized Schmidt decomposition (\ref{ec5}).
After some algebra, we finally arrive at
\begin{eqnarray}
\label{ec15}
\dot{P}&=&2\sqrt{P(1-P)}\ \frac{{\rm Im}
\left(\langle\psi_A\otimes\psi_B,H(\psi_A^\perp\otimes\psi_B^\perp)\rangle\right)
}{\langle\Psi,\Psi\rangle}\ .
\end{eqnarray}
This equation formally agrees with the eq.\ (5) in \citep{dur1}, however,
the definition of the scalar product involved differs from it, of course.\\

\section{Discussion and conclusions}

We can now start our analysis by considering in eq.\ (\ref{ec15})
non-entangled spinors (i.e., $P = 1$). We see immediately that 
(pseudo-) conformal transformations do not change the 
intrinsic entanglement of such spinors. For entangled spinors 
($P\not= 1$) the situation looks differently.
Choosing besides $\gamma_4$ (see eq.\ (\ref{ec7})) 
\begin{eqnarray}
\label{ec16a}
\gamma_1&=&i \sigma_3\otimes\sigma_1\ ,\ \
\gamma_2\ =\ i \sigma_3\otimes\sigma_2\ ,\ \
\gamma_3\ =\ i \sigma_2\otimes\openone_2\ ,\ \ \
\end{eqnarray}
entailing $\gamma_5 = i \sigma_1\otimes\openone_2$
($\sigma_k$ are the standard Pauli matrices, 
$\openone_2$ is the $2\times 2$ unit matrix.),
we can calculate the explicit form of the generators $H$ of 
(pseudo-) conformal transformations (\ref{ec8a})-(\ref{ec8d}).
From eq.\ (\ref{ec15}),
we then find that $M_{12}$, $M_{14}$, $M_{24}$, $D$, $P_3$, and $K_3$
have a vanishing entanglement capability while all 
other (pseudo-) conformal transformations lead to
a change in the intrinsic entanglement of a (entangled) spinor. Recalling the
two-qubit analogue of four-component spinors, these six 
spin representations of
space-time transformations stand in correspondence to the local
(three-parametric) unitary transformations of two qubits which also
do not change their entanglement.
Before attempting to ask any physical questions now it seems to be advisable
to find out what can be said over the entanglement capabilites of 
(pseudo-) conformal transformations if we apply a different,
equally legitimate representation
of the gamma matrices than we just did. Let us choose (by cyclically
permuting the representation matrices for $\gamma_k$, $k = 1,2,3$)
\begin{eqnarray}
\label{ec17}
\gamma_1&=&i \sigma_2\otimes\openone_2\ ,\ \ 
\gamma_2\ =\ i \sigma_3\otimes\sigma_1\ ,\ \
\gamma_3\ =\ i \sigma_3\otimes\sigma_2\ ,\ \ \ 
\end{eqnarray}
entailing again $\gamma_5 = i \sigma_1\otimes\openone_2$.
One then finds a different result: The entanglement capabilites
of the following six generators of (pseudo-) conformal transformations
vanish: $M_{23}$, $M_{24}$, $M_{34}$, $D$, $P_1$, $K_1$. All other 
generators have a non-vanishing entanglement capability within this
gamma matrix representation. In other words, the introduction of a
tensor product structure (TPS) for four-component spinors for which those
are entangled leads to an unphysical (gamma matrix 
representation dependent) result. Furthermore, one might
also like to argue that (passive)
Lorentz/Poincar\'e transformations should not change
the information content of spinors which is related to their intrinsic
entanglement (if it is assumed to have a physical meaning). 
Consequently, we are led to conclude
that only such TPS can be introduced for which a given spinor is
not entangled. It seems worth mentioning that this result justifies
the {\it ad hoc} ansatz (13.4) of Tokuoka 
(\citep{toku1}, Part II, p.\ 162). Concerning the direct product 
representation of solutions of the Dirac equation (\ref{ec1}),
the interested reader is referred to ref.\ \cite{usch1}. In the case of
massless fermions (neutrinos), direct product representations for 
spinors related to the two possible chiralities (\ref{ec1b}), (\ref{ec1c})
are given in eq.\ (3.64), p.\ 31, of \cite{usch1}. Direct product 
representations of spinors related to massive fermions
can be found in eqs.\ (3.42), (3.43), p.\ 27, of \cite{usch1}.\\ 

We conclude this section with some comments on papers which have
been released after the first version of the present article has
been available on the arXiv.  Recently, a paper which 
(as the present article)
relies on Dirac spinors and the Dirac equation in studing 
relativistic entanglement problems has been published
by Debarba and Vianna \cite{deba1}. It should be mentioned that
certain assertions in \cite{deba1} do not conform to the results
obtained in the present article. Specifically, after mentioning
the invariance of the purity ${\rm tr}\;\rho^2$ of the 
density matrix $\rho$ (up to normalization 
our eq.\ (\ref{ec10b})) under Lorentz
transformations Debarba and Vianna
conclude on p.\ 12300003-9 of \cite{deba1} (between eqs.\ (46) 
and (47)): ``A corollary of this straightforward result is the 
invariance of entanglement of pure states under Lorentz
transformations, for in this case the purity of the marginals
characterize the entanglement.''. In other words, 
Debarba and Vianna assert that the entanglement capability of the 
spin representation of Lorentz transformations is always zero
(in contradiction to the above results). Debarba and Vianna do not 
provide the reader of their article with
any line of reasoning for this corollary (Also following an
email request of the author of the present article to the authors
of \cite{deba1} for clarification, no further 
argument/clarification has been given.).
In another paper, Caban, Rembieli\'nski, and W\l odarczyk \cite{caba4} 
have recently discussed a connection between the finite-dimensional, 
non-unitary representation of the Poincar\'e group we are using 
(cf.\ sec.\ \ref{spin}) and some infinite-dimensional,
unitary representation(s) of it. It will be interesting to see
in the future whether the results of the present investigation 
can be useful for the study of entanglement problems on the 
basis of unitary (infinite-dimensional) representations of the 
Poincar\'e group.\\

\section{Final remarks}

The present explorative study based on the formal 
mathematical analogy of four-component
spinors with the pure states of a two-qubit system 
leaves many questions open for further research. 
Spinors (and the Dirac equation) provide us with a
link between space-time (symmetries) and entanglement properties 
of spinors. This does not only apply to (3+1)-dimensional 
Minkowski space-time but also to spinors in higher-dimensional space-times
(a subject which we have not explored in the present article). 
This way entanglement properties of spinors have possibly (at least in
principle) a relation to the problem of space-time dimensionality,
a subject worth of further study.\\

\section*{Acknowledgements\phantomsection
\addcontentsline{toc}{section}{Acknowledgements}}

The author gratefully acknowledges kind hospitality at the 
Theoretical Particle Physics Group of the Department of Physics
and Astronomy of the Vrije Universiteit Amsterdam.\\


\begin{thebibliography}{11}
\bibitem[()]{foot3}
$SU^{\displaystyle\ast}(4)$ is the noncompact version of $SU(4)$ related
to it by the ``unitary trick'' of Weyl \citep{gilm1}.
\end{thebibliography}
\end{document}